# Quantum-like Tests for Contextual Querying

Zeno Toffano and Bich-Liên Doan

SUPELEC, 3 rue Joliot Curie, F-91192, Gif-sur-Yvette, France
`zeno.toffano@supelec.fr, bich-lien.doan@supelec.fr`

**Abstract.** Tests are essential in Information Retrieval (IR), in order to evaluate the effectiveness of a query. Tests intended to exhibit the sense of words in context were undertaken and linked with Quantum Mechanics (QM). Poll tests were undertaken on heterogeneous media such as music and polysemy in foreign languages. Interference effects are shown in the results. Bell inequality was used leading to a significant spread in the results of the poll tests but without violating the classical limit. Then an automatic pertinence measure tool on texts has been developed using the HAL algorithm using an orthonormal vector decomposition model. In this case the spread in the values can lead to the violation of the Bell inequality even beyond Cirel'son bound.

## 1  Introduction

In this work we emphasize an effort to develop original "quantum-like" tests that could be useful to the domain of Information Retrieval (IR).

IR systems are evaluated by human assessment. Standard test collections are traditionally used to calculate performance but the scores are difficult to interpret due to the high variability of queries (query difficulty). Ambiguity of natural language makes the issue of the effectiveness (precision measure) of IR systems both essential and contested. This shows that IR remains an empirical and subjective discipline, even with a sophisticated suite of mathematical models [1].

Intuitively, it is clear that the same request made by two different users at two different times may reflect different expectations of outcome. One way to build a context is to log the behavior of the user, or to take into account the environment of the user (profile, interactions) as parameters in the query process. For example, Google already uses the history of queries, especially for spell checking or expansion of the query. That is why contextual meaning is linked with ambiguity and with the sense of words. Ambiguity is mainly caused by polysemy (one term with different meanings) and synonymy (different words for the same meaning). The work of Krovets & Croft [2] shows that there is considerable ambiguity even in a specialized database. Recent research tried to classify these ambiguities and to construct test collections [3] in or-der to measure their impact on IR systems. Context is used to disambiguate terms, and Melucci [4] showed that a query and a document can be generalized in different contexts as vectors, where the likelihood of context of a set of documents is considered. Quantum Mechanics (QM) has been invoked to enrich the search capabilities in IR by Rijsbergen [5] by using the mathematical formalism of the Hilbert vector space. The strong point is that both QM and IR deal with the interaction between a user (the observer in QM) and the object under observation. The probabilistic aspects are central in all attempts to link QM with IR. Consensus is observed for the role of conditional probabilities associated with interference effects leading to some contextual measure (mental, subjective character) when combining several entities or tests [6].

Analogies between concepts issued from QM with IR tools and objects have been made by several authors. Widdows [7] uses the quantum formalism for its practical advantages and Arafat [8] shows that user needs can be represented by a state vector. Other analogies have been stated by Li & Cunnigham [9] such as: state vector/objects in a collection; observable/query; eigenvalues/relevant or not for one object; probability of getting one eigenvalue/relevance degree of object to query.

In this paper, we will present different tests and some IR models which were developed in our Faculty. Experimental tests aim to show the quantum-like character particularly focusing on the polysemy of terms, the correlation of the sense of words and the query formulation using interference and also a Bell inequality test.

## 2      Test Framework

It is important to remember essential concepts considered in Quantum Mechanics which can be associated with IR objects.

- **State Vector**: defines a physical state (particles but not only…). Different state vectors can be superposed leading to other state vectors.
- **Outcomes of a measurement** are the eigenvalues of the corresponding measuring operator, the "observable". After the measurement the state "becomes" the corresponding eigenvector.
- **Probability** to obtain a given result is the square module of the dot product of the initial state vector by the state vector immediately after measurement. This works because state-vectors are orthonormalized.
- **Complementarity or Incompatibility** : measurements are "incompatible" if the measurement order can lead to different final results. Heisenberg's uncertainty principle can be deduced from this property.
- **Coherence and Interference**: QM operates in a complex space with the important role of "phase" responsible of "coherence" and "interference"
- **Entanglement**: is a particular combination (tensor product) of two or more quantum states. "Bell Inequalities" are violated in these states.

In this approach one has to preserve "quantum-like behavior" [10] possibly by avoiding early inferring. The "Incompatibility Advantage" could bring context-related results, for example using complementary media by mixing texts, images [11], video, music or using different languages.

Investigating in new test validation protocols is an interesting line of research, for example by the application of the Bell inequalities. This can be done by calculating the Bell parameter $S_{Bell}$ for the outcomes of two mutual tests, for definition see Eq. (3) hereafter. $S_{Bell}$ between 0 and 2 corresponds to the classical local case. $2 \leq S_{Bell} \leq 2\sqrt{2}$ corresponds to quantum nonlocal entangled states, i.e. obtained experimentally with photon polarizations [12]. $S_{Bell} > 2\sqrt{2}$ is what is now known as the Cirel'son bound and goes beyond quantum correlations [13].

Some authors lay the possibility that human language does not in all cases comply with the classical Bell inequality [14], especially for ambiguous words whose meaning depends upon their context of use. Here we adopt an experimental approach, without formalization in order to discover some analogies between quantum-like measures and IR relevance measures.

The following pilot tests were conducted by three groups of students. These tests were implemented under different conditions using different media (text, sound), and different languages. Tested subjects had different backgrounds, and they were asked questions without knowledge of the results of the previous tests, or conversely with the knowledge of previous answers. Tests were made on French speaking people (test 1 on music and test 3 fruit/vegetable) and one on Chinese speaking people (test 2 on foreign languages). Results are presented in their original form.

## 3    Pilot Tests Based on Polls

### 3.1    A Pilot Test for Heterogeneous Media : Music

| Music Excerpts | Rock | Blues | Rock + Blues |
|---|---|---|---|
| « Dazed and confused » | 5.5 | 4.5 | 10 |
| « Susie Q » | 8 | 2.5 | 10.5 |
| « That's all right » | 4.5 | 5.75 | 10.25 |
| « Folsom prison blues » | 4 | 4 | 8 |
| « The wind cries Mary » | 4.75 | 6.25 | 11 |
| « Don't let me down » | 8.25 | 3.25 | 11 |
| « Tenth avenue freeze out » | 3.5 | 3.75 | 7.25 |
| « Since I've been loving you » | 2 | 8 | 10 |
| « I heard it through the grapevine» | 7.5 | 3.25 | 10.75 |

**Table 1.** Musical excerpts question Rock or Blues average for 4 respondents.

The first test measured correlations between concepts from a medium other than language. The test proposes nine musical excerpts. Four subjects were asked to rate, on a scale from 0 to 10, whether these excerpts fall under the category "rock" or "blues". 4 persons were interviewed. The score is an average of four scores given by the respondents. In Table 1 the sum of the results for both categories, appearing in the right column, is very rarely equal to 10 (two occurrences) indicating that the chosen categories are certainly not mutually exclusive. Here we estimate the correlation between a musical excerpt and a word ("rock", "blues"). This could be interpreted as some kind of interference between concepts of different media.

### 3.2 A Pilot Test in Heterogeneous Media : Foreign Language

The second pilot test was carried out in two experiments at once: it assessed correlations in a foreign language (Chinese here). It aimed to show the role of polysemy of words. The question was to quantify the correlation with different meanings of the proposed words. 4 persons were interviewed (all Chinese).

Results are shown on Table 2: the four answers are proposed without averaging. The great dispersion of results shows that the polysemy of words is influential in this correlative test.

| chinese word | Polysemy | Scores |
|---|---|---|
| 笔记本 | *ordinateur portable*  (laptop computer) | 9 9 6 4 |
|  | *carnet*  (paper notebook) | 4 6 6 7 |
| 性 | *sexe*  (sex) | 9 9 5 6 |
|  | *caractère*  (character) | 2 6 6 5 |
| 生 | *vie*  (life) | 5 5 8 5 |
|  | *naître*  (be born) | 9 9 6 7 |
| 清 | *dynastie des Qing*  (Qing dynasty) | 8 6 4 8 |
|  | *juste et intègre*  (fair end honest) | 2 5 6 3 |
| 出入 | *sortir et entrer*  (go in and out) | 8 6 6 6 |
|  | *le fait de ne pas s'accorder*  (the failure to agree) | 4 6 9 6 |

**Table 2.** Polysemy test for Chinese words for 4 respondents.

### 3.3 Correlation Tests on Words

We present here a pilot test inspired from the work of Aerts [15] using an extension in order to obtain Bell Inequality test scores. The following experiment is undertaken: respondents give a score depending on the degree of relationship of the pro-posed word to a given category. A word such as "tomato" which has no clear belonging to a specific category (fruit or vegetable) shows interesting features. We can observe the phenomenon of overextension, meaning that its relationship in both categories is stronger than individually.

In analogy with QM we can write the corresponding category state  $|Fruit\rangle$

The test result is represented by the average $\mu$ of the responses to the question "... $appartient$ à ..." ("… belongs to"). In the case of category $Fruit$ and object $Tomate$ we can write the following expression in Eq. 1 corresponding to the mean value in the QM form (question/operator in the center and the state at the right, ket, and at the left, conjugate bra) :

$$\mu(fruit) = \langle fruit|"Tomate\ appartient\ \text{à} ..."|fruit\rangle \quad (1)$$

In order to take into account the effect of context, categories $A$ and $B$ are represented by three vectors, $|A\rangle$, $|B\rangle$ and $|A\ or\ B\rangle$. From this model we can show interference. The result will be in the form:

$$\mu(A\ or\ B) = (\mu(A) + \mu(B))/2 + c\sqrt{\mu(A)\mu(B)} \cos\phi \quad \text{with} \quad c \geq 0 \quad (2)$$

In Eq. 2 the variation of the phase angle $\phi$ can therefore achieve a greater range of values for $\mu(A\ or\ B)$ than for the individual averages $\mu(A)$ and $\mu(B)$.

We chose not to interview the same persons on different categories (fruits, vegetables and fruits or vegetables) so that their responses would not be influenced by the context of other categories. Thus, persons who we asked "$l'ail\ est-il\ un\ Fruit\ pour\ vous?$" ("is garlic a fruit for you?") did not know that the other two possible categories were $Légume$ (vegetable) and $Fruit\ ou\ Légume$. $\mu(F)$ corresponds to the average of responses (the maximum score is 1) for the object belonging to the category $Fruit$ , $\mu(L)$ for $Légume$ and $\mu(F\ or\ L)$ for belonging to both.

We calculated the Bell parameter $S_{Bell}$ given in Eq. 3 as a function of the mean expectation values $E$ of correlated tests with two valued results for example $u_F$ "belongs to $Fruit$ " or $u'_F$ "does not belong to $Fruit$".

$$S_{Bell}(F,L) = E(u_F, u_L) + E(u_F, u'_L) + E(u'_F, u'_L) - E(u'_F, u_L) \quad (3)$$

The explicit Bell parameter is calculated in Eq. 4 using the expectation values function of the test averages $\mu$ :

$$S_{Bell}^{test} = \mu(F)\mu(L) + \mu(F)(10 - \mu(L)) + |(10 - \mu(F\ or\ L)) + (10 - \mu(F))\mu(L)| \quad (4)$$

The scores in Table 3 range between 0 and 1. The number of respondents was 29. First, we note that the probability $\mu(F\ or\ L)$ does not always lie between the probabilities $\mu(F)$ and $\mu(L)$. There are cases of sub-extension for example with *Poivre* (Pepper) but also of over-extension with *Amande* (Almond).

In our test $S_{Bell}$ never reaches the value 2 in absolute value and so does not violate the Bell inequality.

| *Objet* (Object) | *μ(Fruit)* | *μ(Légume)* | *μ(F ou L)* | $S_{Bell}$ |
|---|---|---|---|---|
| *ail* (garlic) | 0.16 | 0.525 | 0.333 | 0.39 |
| *amande* (almond) | 0.68 | 0.075 | 0.833 | 0.82 |
| *betterave* (beet) | 0.2 | 0.4 | 0.533 | 0.35 |
| *brocolis* (broccoli) | 0.04 | 0.925 | 1 | 0.93 |
| *champignon* (mushroom) | 0.06 | 0.75 | 0.6 | 0.36 |
| *chou-fleur* (cauliflower) | 0.06 | 0.925 | 0.933 | 0.86 |
| *concombre* (cucumber) | 0.18 | 0.725 | 0.833 | 0.61 |
| *cornichon* (gherkin) | 0.26 | 0.475 | 0.8 | 0.41 |
| *épinards* (spinach) | 0.04 | 0.95 | 0.8 | 0.75 |
| *haricot* (bean) | 0.1 | 0.975 | 0.866 | 0.84 |
| *noix de coco* (coco nut) | 0.96 | 0.025 | 0.6 | 1.36 |
| *olive* (olive) | 0.6 | 0.4 | 0.666 | 0.77 |
| *persil* (parsley) | 0.14 | 0.35 | 0.466 | 0.37 |
| *poivre* (pepper) | 0.08 | 0.05 | 0 | 1.034 |
| *pomme de terre* (potato) | 0.16 | 0.45 | 0.166 | 0.62 |
| *pomme* (apple) | 0.94 | 0.025 | 1 | 0.94 |
| *gingembre*…(ginger) | 0.1 | 0.225 | 0.266 | 0.63 |
| *raisin* (grapes) | 1 | 0.025 | 1 | 1 |
| *tomate*…(tomato) | 0.8 | 0.6 | 1 | 0.92 |

**Table 3.** Test on the belonging on two categories for 29 respondents.

In the next section we will present another test using the HAL algorithm for which we will get Bell Inequality parameters greater than 2.

## 4 Model and IR Implemented Tests Using HAL Algorithm

Our goal here is to classify a set of texts using the HAL algorithm (Hyperspace Analog to Language) following a search criteria that depends on the relation to a keyword [16]. Given a text consisting of a vocabulary of $n$ words, HAL returns a $n \times n$ matrix constructed by moving a window of length $l$ on a given text. Punctuation and paragraphs are ignored. Consider a set of $n$ documents $T_1, \ldots T_n$ based on a finite vocabulary $V = \{e_i\}_{i \in I}$. In our approach to link QM with RI we consider that the words ($v_i$) form an orthogonal basis and represent coordinates according to our measurement criteria (semantic vectors).

For every text $T_j$ we consider the complete set of distinct words which compose the text : $V_j = \{v_k\}_{k \in I_j}$. Some words, in the text, possess a contextual meaning. So we will differentiate words ($v_k$) with a contextual meaning from the "raw" ($e_i$) words without context. We will consider the projection of a "contextual" word ($v_k$) on the basis ($e_i$) (by completing with zeros the base vectors $e_i$ not present in the text).

We use the HAL algorithm, without considering the order of words in the text, then we create a symmetrical interaction matrix $H$. The element $H[i,j]$ reflects the weight of the combination of word $i$, taken in the context of word $j$. The greater the weight of a word, the more often it appears in the context of the other word. The program will match a single semantic term for a word in singular or plural form.

Two types of query have been undertaken: "$A\ AND\ B$", which catches a word $A$ in the sense of $B$ and "$B\ WITHOUT\ A$", which catches the word $A$ without the sense of $B$.

- $e_i\ AND\ e_j$ : for each text $T_k$, consider $v_i$ and $v_j$ the words corresponding to the base vectors $e_i$ and $e_j$. We define the scalar product of vector $v_i$ corresponding to $H_k[i]$ with vector $v_j$ corresponding to $H_k[j]$ by :

$$< H_k[i] | H_k[j] > = \sum_{l=1\ldots n} H_k[l][i] * H_k[l][j] \qquad (5)$$

This expression corresponds to the sum of products of the weights of common terms and the product of the cross weights. Texts are then measured according to this score.

- $e_i\ WITHOUT\ e_j$ : for each text $T_k$, consider $v_i$ and $v_j$ as before. We define the new semantic vector:

$$X_k = H_k[i] - (H_k[i] | H_k[j]) * H_k[j] \qquad (6)$$

Eq. (6) corresponds to the projection of the semantic vector on the complement of the word to eliminate. We then take the square norm of this vector as the score.

By applying HAL to a text, because the algorithm considers only terms of both sides of $m$, we find that the weight of a word $m$ compared to itself is zero (or low if the same word appears in the window). Now, the semantics associated with vector $m$ should have its largest component along the word $m$. We therefore encompass the window of HAL, taking now the value $l + 1$ (if the window has length $l$).

Tests are undertaken in French language (French words in italics). We consider several texts referring to the word tomato ($Tomate$) that may be considered either as a fruit ($Fruit$) or as a vegetable ($Légume$). Documents were selected from the Internet (Google) using keywords such as text, descriptions and articles [17], [18], [19]. We have not considered the origin of these documents thus leaving a random generation character.

The results show that the program sorts out the texts in a realistic order, both for research of operators AND (*ET*) and WITHOUT (*SANS*). The scores we get for the three documents are shown on Table 4.

We wanted to take into account interactions so we have refined the classification : we consider simple words strongly related to semantic vectors $Tomate$ and $Légume$, e.g. $Plante$ (Plant) in the previous case example. To do this we use the scalar product, which is the product of the 2x2 contextual vector components $Tomate$ and $Légume$. This represents a sort of contextual entanglement of these two vectors.

In the query using exclusion, for example $Tomate\ SANS\ Légume$ (tomato WITHOUT vegetable)," we do not only remove the component from the contextual vector $Tomate$ along the simple word $Légume$ (vegetable), we also reduce all components involved in the context vector $Légume$ and also for example $Plante$. We weight the reduction by the scalar product, which represents the correlation of the two words.

We decided to use this search engine to perform tests on Bell's inequality. We consider four words: $Tomate$ (Tomato), $Légume$ (Vegetable), $Fruit$ (Fruit) et $Plante$ (Plant). For each text we apply our algorithm to obtain the score of the following $ET$ (AND) queries: $Tomate\ ET\ Légume$, $Tomate\ ET\ Fruit$, $Plante\ ET\ Légume$ and $Plante\ ET\ Fruit$. The test is as follows: if we consider a text in a corpus of other unknown texts, and undertake the search described above, will this text end up in the first position in the search? If the answer is yes, we attribute the result $+1$, otherwise $-1$. The score given by our algorithm $x$ corresponds to the probability for this text to be first in the request. We can consider that the maximum score of other requests can be represented by random variable equally distributed on [0,1], this case corresponds to a large number of requests on arbitrary documents. Thus the average value of the test is $E = 2x - 1$.

Next, we determine the Bell parameter as follows:

$$S_{Bell}^{HAL} = |E(x_{TF}) - E(x_{TL})| + |E(x_{PF}) - E(x_{PL})| \quad \text{with} \quad E(x) = 2x - 1 \quad (7)$$

where $E(x_{AB})$ is the mean expectation value derived for the search "$A\ ET\ B$" ("$A\ AND\ B$")

The results are shown on Table 4. In one cases the Bell parameter $S_{Bell}$ calculated from Eq. 7 exceeds the classical limit of 2 but is below $2\sqrt{2}$ this cas corresponds to quantum entanglement.

| Document n° | 1 | 2 | 3 | | |
|---|---|---|---|---|---|
| *Tomate* ET *Fruit* | 0.788 | 0.581 | 0.373 | **score** | $x_{TF}$ |
| *Tomate* ET *Légume* | 0.349 | 0.469 | 0.213 | **score** | $x_{TL}$ |
| *Plante* ET *Fruit* | 0.651 | 0 | 0.385 | **score** | $x_{PF}$ |
| *Plante* ET *Légume* | 0.315 | 0 | 0.223 | **score** | $x_{PL}$ |
| | 0.947 | 2.23 | 1.105 | **Bell param.** *S* | |

**Table 4.** Bell parameter tests for correlations on words using the HAL algorithm.

In the next example $S_{Bell}$ calculated from Eq. 7 exceeds the quantum entanglement limit $S_{Bell} = 2\sqrt{2}$ (Cirel'son bound). This case is illustrated on Table 5 with the requests *Tomate ET Fruit*, *Tomate ET Utilisé* (Tomato AND Used), *Voiture ET Fruit* (Car AND Fruit), *Voiture ET Utilisé* (Used AND Car). Although this research is somewhat unrealistic, it illustrates non-pathological cases (we do not obtain {1, 0, 0, 0}).

| Document n° | 1 | | |
|---|---|---|---|
| *Tomate* ET *Fruit* | 0.788 | **score** | $x_{TF}$ |
| *Tomate* ET *Utilisé* | 0.091 | **score** | $x_{TU}$ |
| *Voiture* ET *Fruit* | 0 | **score** | $x_{VF}$ |
| *Voiture* ET *Utilisé* | 0 | **score** | $x_{VU}$ |
| | 3.384 | **Bell parameter** *S* | |

**Table 5.** Bell parameter test beyond Cirel'son bound.

In our model we constructed vectors which decompose on semantic word-bases, similar to the bases of eigenvectors in quantum mechanics. By using our search engine, we could design a Bell-type test. Our test allowed us to obtain a Bell parameter $S_{Bell}$ taking values between 0 and 4. In some cases Bell's inequality is violated : $S_{Bell}^{HAL} = 2.23$ in Document n°2, Table 4 and $S_{Bell}^{HAL} = 3.384$ in Table 5.

Even if the analogy with quantum phenomena must be made very carefully we consider that we made a small step forward from previous discussions [15], in the sense that our results are not always 0, 2 or 4 due to binary test results $+1$ or $-1$. One reason is because our model is not deterministic using a measure corresponding to a physical distance between words in a text.

To confirm these results, the engine should be tested on a much larger database.

## 5   Conclusion, Perspectives and Discussion

We carried out tests on different groups of persons and heterogeneous IR objects to obtain a quantitative evaluation of the relevance of information in different contexts. Test were undertaken under different conditions:
- Pilot tests on heterogeneous media and languages.
- IR model and automatized test using the HAL algorithm.

The experiments demonstrate some analogies between quantum-like measures and IR relevance measures. These preliminary results are promising for further research on more formal indicators in order to evaluate IR systems inspired on QM concepts.

To overcome the current limitation, future work will consist in :
- Increasing data volume tests and testers.
- Classifying test data according to the specific conditions (heterogeneity of observed items, knowledge of users, temporal aspects).
- Proposing a formal framework test in order to go further in the analogies between QM and relevance measures.
- Providing an argued interpretation of results of tests based on Bell inequalities in the case of the pilot tests and when using the HAL procedure.

We strongly believe that QM can lead to benefits in the IR research field. But the motivations of these domains are very different. Mutual understanding needs much humility and efforts in order to make concepts comprehensive for both fields. In recent years much has been expected from quantum approaches outside Physics. But in some cases, due to the lack of results, this has led to disappointment, perhaps due to naive behaviors and uncritical fascination on behalf of QM.

A frequent argument is that actually only the mathematical framework of QM is used. The physicist community observes suspiciously attempts to widen the domain of application of QM. Some debates in the community are very old and touchy for example on the classic/quantum frontier, on non-locality, on wave-particle duality, on the Schrödinger cat paradox. But we want to stress that a continuously increasing number of physicists and also engineers are confronted to quantum-like applications. Today quantum concepts are omnipresent even at a macroscopic scale in the domains of applied physics and engineering, such as optoelectronics and lasers, semiconductors, radio waves... So we think that new concepts and analogies can also be found in these applied fields.

The initiators of this work are with an Engineering Faculty *Grande École* SUPELEC mostly centered on Information Sciences and Electrical Engineering: one is a Physicist working in the field of optoelectronics and lecturing Quantum Mechanics; the other has been working since several years in the field of Information Retrieval. Different backgrounds can be an advantage for this kind of multidisciplinary and collaborative work.

## 6      Acknowledgements

We would like to associate to this paper and thank the groups of students which participated to the test definition and exploitation: Alain SIROIS and Yuhui XIONG for the pilot test poll on music and foreign language, Felix BEZIZ and Charlotte RUDNIK for the pilot poll test on fruit/vegetable and Olivier BALLAND and Alexandre MULLER for the implementation and test using the HAL algorithm.

## 7      References

1. Manning, C. D., Raghavan, P., Schütze, H.: Introduction to Information Retrieval, Cambridge University Press. (2008).
2. Krovets, R., Croft, B. W.: Lexical Ambiguity and information Retrieval. ACM Transactions on Information Systems, 10(2), pp.115-141, (1992)
3. Sanderson, M.: Word-sense Disambiguation and Information Retrieval. In proceedings of ACM-SIGIR, (1994)
4. Melucci, M.: Context Modeling and Discovery Using Vector Space Bases. In: Proc. of the ACM Conf. on Inf. and Knowledge Management, pp. 808–815 (2005)
5. van Rijsbergen, C.J.: The Geometry of Information Retrieval. Cambridge University Press, Cambridge (2004)
6. Buchanan, M.: Q. minds: Why we think like quarks, New Scient., 2828, pp. 34-37 (2011)
7. Widdows, D: Semantic vector products : Some initial investigations, In Proc. of the Second AAAI Symposium on Quantum Interaction, volume 26, (2008)
8. Arafat, S.: Foundations research in information retrieval inspired by quantum theory. PhD thesis, University of Glasgow (2008)
9. Li Y., Cunningham H.: Geometric and Quantum Methods for Information Retrieval , SIGIR Forum, , 42, pp.22-32 (2008)
10. Asano, M., Ohya1, M., Tanaka, Y., Yamato, I., Basieva, I., Khrennikov, A. : Quantum-Like Paradigm: From Molecular Biology to Cognitive Psychology, QI 2011, LNCS 7052, Springer-Verlag , pp. 182–191, (2011)
11. Kaliciak, L., Wang, J., Song, D., Zhang, P., Hou, Y.: Contextual Image Annotation via Projection and Quantum Theory Inspired Measurement for Integration of Text and Visual Features, QI 2011, LNCS 7052, Springer-Verlag, pp. 217–222, (2011)
12. Rössler O. E., Parisi, J., Fröhlich, D., Toffano, Z.: Direction-Correlated Correlated Photons Cannot Self-Interfere: A Prediction, Z. Naturforsch., 61a, 418 (2006)
13. Cabello, A.: Violating Bell's inequality beyond Cirel'son's bound, Physical Review Letters, vol. 88, no. 6, 060403, (2002)
14. Kitto, K., Ramm, B., Bruza, P.: Testing for the Non-Separability of Bi-Ambiguous Compounds, Proceedings of the AAAI Fall Symposium on Quantum Informatics for Cognitive, Social, and Semantic Processes (QI 2010) pp. 62-69, (2010)
15. Aerts, D.: Q. Structure in cognition, J. Of Math. Psychology, 53, pp. 314-348 (2009)
16. Lund, K., Burgess, C.: Producing high-dimensional semantic spaces from lexical co-occurrence. Behav Research Methods Instruments and Computers 28, pp. 203-208, (1996)
17. http://www.pourquois.com/nature_bio/pourquoi-tomates-sont-fruits-pas-legumes.html
18. http://www.evene.fr/forum/reponse.php?id_discussion=7390&id_message=183645&hunk=1&msidx=2
19. http://ventre-plat-tip.blogspot.fr/2011/11/11-faits-amusants-sur-les-tomates.html